\documentstyle[aaspp4]{article}
 
\lefthead{Elmegreen}
\righthead{A Prediction of Brown Dwarfs}
\slugcomment{{\it Astrophysical Journal}, Vol 522, September 10, 1999}
\begin{document}
 
\title{A Prediction of Brown Dwarfs in Ultracold Molecular Gas}
 
\author{Bruce G.~Elmegreen\altaffilmark{1}}
\altaffiltext{1}{IBM Research Division, T.J. Watson Research Center,
P.O. Box 218, Yorktown Heights, NY 10598, bge@watson.ibm.com}
 
\begin{abstract} A recent model for the stellar initial mass function
(IMF), in which the stellar masses are randomly sampled down to the
thermal Jeans mass from hierarchically structured pre-stellar clouds,
predicts that regions of ultra-cold CO gas, such as those recently found in
nearby galaxies by Allen and collaborators, should make an abundance of
Brown Dwarfs with relatively few normal stars. This result comes from
the low value of the thermal Jeans mass, which scales as $M_J\propto
T^2/P^{1/2}$ for temperature $T$ and pressure $P$, considering that the
hierarchical cloud model always gives the Salpeter IMF slope above this
lower mass limit. The ultracold CO clouds in the inner disk of M31 have
$T\sim3$K and pressures that are probably $10\times$ higher than in the
solar neighborhood. This gives a mass at the peak of the IMF equal to
$0.01$ M$_\odot$, well below the Brown Dwarf limit of $0.08$ M$_\odot$.
Using a functional approximation to the IMF given by
$\left(1-e^{-\left[M/M_J\right]^2}\right) M^{-1.35}d\log M$ for $M>M_J$,
which fits the local IMF for the expected value of
$M_J\sim0.3$ M$_\odot$, an IMF
with $M_J=0.01$ M$_\odot$ in M31 has 50\% of the mass and 90\% of the
objects below the Brown Dwarf limit. The brightest of the Brown Dwarfs
in M31 should have an apparent, extinction-corrected K-band magnitude of
$\sim21$ mag in their pre-main sequence phase. For typical
star-formation efficiencies of $\le10$\%, Brown Dwarfs and any
associated stars up to $\sim2.5$ M$_\odot$ should not heat the gas
noticeably, but if the IMF continues up to arbitrarily high masses, then
the star formation efficiency has to be $\le10^{-4}$ to avoid heating
from massive stars. \end{abstract}

keywords: stars: brown dwarfs --- stars: mass function --- 
galaxies: stellar content

\section{Introduction}

The recent observation of ultracold CO in the inner disk of M31 by Allen
et al. (1995), and the inference that ultracold CO exists in some spiral
arm dustlanes where neither CO nor HI have been detected in spite of a
large dust extinction (M83: Allen, Atherton \& Tilanus 1985; Tilanus \&
Allen 1993; M51: Tilanus \& Allen 1989; M100: Rand 1995; M81: Allen et
al. 1997) lead us to wonder about the form of the initial mass function
(IMF) for any stars that are born at such low temperatures. Here we
propose that star formation in ultracold gas is biased towards Brown
Dwarfs, and we show that these objects, along with the stars that are
likely to form with them, could have escaped detection up to now because
they do not heat the gas enough to raise the temperature to normal
levels, nor do they appear as stellar objects in existing images. 

In the Solar neighborhood, Brown Dwarfs are so uncommon that the IMF
cannot continue to rise below about $0.1$ M$_\odot$ (Zuckerman \&
Becklin 1992; Pound \& Blitz 1993, 1995; Reid \& Gazis 1997a,b).  The
number of possible Brown Dwarfs is consistent with an extension of the
IMF flattening seen between $\sim1$ M$_\odot$ and 0.4 M$_\odot$ (Reid
1998).  The relative number of Brown Dwarfs could be much higher in the
ultracold gas of M31, however. The detection of numerous Brown Dwarfs
there would be an important confirmation of recent IMF models.

\section{The Case for Uniformity in the IMF}

The IMF is observed to be remarkably uniform from region to region in
any one galaxy (see reviews in Massey 1998; Elmegreen 1999a), covering
star formation that spans a factor of $\sim200$ in density (Massey \&
Hunter 1998; Luhman \& Rieke 1998) and a factor of $\sim10$ in
metallicity (Freedman 1985; Massey, Johnson \& DeGioia-Eastwood 1995).
It is also about the same in different HII regions in various galaxies
(Bresolin \& Kennicutt 1997), and in many different galaxies on average
(Kennicutt, Tamblyn \& Congdon 1994), as indicated by the equivalent
widths of hydrogen emission lines. Detailed studies of color magnitude
diagrams in the LMC and local dwarf galaxies give the same IMF too
(Greggio et al. 1993; Marconi et al. 1995; Holtzman et al. 1997;
Grillmair et al. 1998). 

There is apparently some uniformity in the IMF with time also, because
stars with a wide range of ages in our Galaxy all have about the same
function, as suggested by halo stars (Nissen et al. 1994), and globular
cluster stars (De Marchi \& Paresce 1997). Similarly, a nearly universal
IMF was found from abundance ratios (e.g., Fe/O, reflecting the ratio of
low mass to high mass supernova processing) in QSO damped Ly$\alpha$ (Lu
et al. 1996) and Ly$\alpha$ forest (Wyse 1998) lines, the intracluster
medium (Renzini et al. 1993; Wyse 1997, 1998; but see Loewenstein
\& Mushotzky 1996), and elliptical galaxies (see review in Wyse 1998).

An IMF biased towards high-mass stars has been suggested for starburst
regions, based on the ratio of luminous to dynamical mass (Rieke et al.
1980, 1993; Kronberg, Biermann, \& Schwab 1985; Wright et al. 1988),
galactic evolution models (Doane \& Matthews 1993), spectroscopic line
ratios (Doyon, Joseph, \& Wright 1994; Smith et al. 1995), and infrared
excesses (Smith, Herter \& Haynes 1998). However, a lower extinction
correction for M82 makes the IMF there normal (Devereux 1989; Satyapal
et al. 1995, 1997), and more recent evolutionary models (Schaerer 1996),
multiwavelength spectroscopy and broad-band infrared photometry
(Calzetti 1997), and emission line spectroscopy (Stasi\'nska \&
Leitherer 1996) give normal IMFs too. Large IMF shifts in starburst
galaxies should also produce unobserved red populations of stars after
the turnoff age reaches the stellar lifetime at the truncation mass
(Charlot et al. 1993), and too high an oxygen abundance (Wang \& Silk
1993).

An IMF shift towards lower mass stars has been reported for
the extreme field by Massey et al. (1995).  However, this
result could also come from a normal IMF in each star-forming region if
massive stars form preferentially in high mass clouds and 
stop further star formation when they do (Elmegreen
1999b). The extreme field IMF cannot be typical, because
the IMFs in most clusters and associations are about the same as the
galaxy-integrated IMFs. 

There are also IMF dips, gaps, and a $\pm0.5$ cluster-to-cluster
variation in the power law slope (Scalo 1998), but such variations
are expected statistically given the small numbers of stars that are
usually included in cluster studies (Elmegreen 1999b). 

There are few theoretical predictions of IMF variations. Fabian (1994)
predicted that the IMF would be biased towards low mass stars at the
cores of galaxy cluster cooling flows because of the low Jeans mass that
results from the expected high pressures there.  Larson (1999) proposed
that a top-heavy IMF in the early Universe could explain the G-dwarf
problem, the high temperature and high metal abundance of intracluster
gas, and the large luminosities of young elliptical galaxies. 
Neither of these predictions have been directly confirmed, however.

In what follows, we discuss the possible dependence of the minimum
stellar mass on cloud temperature and pressure. First we discuss
the minimum mass on general terms, and then we apply the
results to ultracold clouds in M31.  

\section{The Thermal Jeans Mass as a Limit to Stellar Mass}

A lower mass limit is required for star formation because cloud pieces
in pre-stellar clouds extend down to masses much smaller than the
smallest stellar mass. Clumps with masses as low as $10^{-4}$ M$_\odot$
have been observed in great abundance as part of the normal, power-law
clump mass function in the Polaris spur (Heithausen et al. 1998; Kramer
et al. 1998). This means that the stellar mass range is only a small
part of the total mass range for cloud clumps, and that there must be
some physical process which limits the stellar mass at the low end. 

One possibility is that the minimum stellar mass is proportional to
the thermal Jeans mass, 
\begin{equation}
M_{J}=0.35\left({{T}\over{10K}}\right)^{2}\left({{P}\over
{10^6\;k_B}}\right)^{-1/2}\;\;{\rm M}_\odot \label{eq:mj} \end{equation}
for temperature $T$ and cloud-core pressure $P$. In this model, cloud
pieces much smaller than $M_J$ are not likely to become stars because
they are not strongly self-gravitating (Larson 1992; Elmegreen 1997, 1999b). 

The mass of a star that actually forms in such a minimum unstable cloud
piece can be smaller than $M_J$, perhaps by a factor of 10, because not
all of the gas goes into stars, and because the resulting stellar
system could be binary, in which case the stars would have to share the
clump mass.  For this reason, $M_J$, or some factor of order unity
times $M_J$, is identified with the mass at the break point in the IMF,
where the power law first becomes flat, not with the minimum mass of a
star.  This is because the model predicts only a fundamental change in
the physical properties of cloud clumps at this mass, but does not
specify how the stars actually form inside the clumps.  Indeed, stellar
masses are often observed to continue down to at least one-quarter of
the mass of the break point, but such stars are not as common as would
be expected from an extrapolation of the Salpeter power-law function.

The expression for $M_J$ is the Bonner-Ebert condition for stability
of a non-magnetic, pressure-bounded isothermal sphere.  Magnetic forces
increase this critical mass (e.g., Mouschovias \& Spitzer 1976), so
they should not be included in the minimum value.  Also, the
mass-to-flux ratio varies in a turbulent cloud as a result of random
compressions and magnetic diffusion, so regions with large mass-to-flux
ratios will develop spontaneously and systematically over time,
suggesting again that equation (\ref{eq:mj}), without magnetic fields, is
the proper condition (see also Nakano 1998).

The numerator in $M_J$ contains the thermal
temperature, rather than the turbulent rms speed, because turbulence
increases the critical mass, and because the smallest stars will
generally form on such small scales that the relative turbulent speed
is less than the sound speed anyway.  The pressure in the expression
for $M_J$ is the total pressure at the boundary of the isothermal
sphere.  This pressure is not well defined for a real cloud because it
comes partly from turbulent interclump motions and partly from 
thermal pressure in the interclump and overlying 
media (e.g.,  Hunter \& Fleck 1982; Ballesteros-Paredes, Vazquez-Semadeni
\& Scalo 1999).  The value of it for any particular cloud is not well
defined either because pressure varies from place to place.
Nevertheless, the total pressure is much more uniform in a cloud than
either the turbulent or the thermal pressures alone, which interchange
roles during transient compressions, and also more uniform than the
density because of the Larson (1981) scaling relations, which make
$\rho \Delta v^2\sim$ constant for density $\rho$ and rms speed $\Delta
v$.  In Taurus-like clouds, $P\sim10^5$ k$_B$, in GMCs, $P\sim10^6$
k$_B$, and in cluster-forming GMC cores, $P\sim10^7$ k$_B$.  

These differences in $P$ suggest that the IMF may differ as well, which
means that the lowest-mass stars might be smaller in GMCs than in
Taurus-like clouds.  Any such differences would not be large, however,
because of the square root dependence of $M_J$ on $P$. Small IMF
differences would not have been recognized either.  Most stars form in
GMCs under similar conditions, so their local IMFs are similar.  Where
the conditions differ, as in Taurus, there are too few stars to get an
accurate IMF. In addition, the expected decrease in the mass at the
peak of the IMF for dense GMC cores is offset by stellar mass
segregation (a concentration of massive stars near the center) and by a
slightly increased temperature from nearby OB stars.

Strong variations in $M_J$ are not expected on a galactic scale
either.  The numerator in $M_J$ is roughly proportional to the cloud
cooling rate and the denominator is roughly proportional to the
background galactic heating rate from stars and cosmic rays (Elmegreen
1997; 1999b).  Thus molecular clouds in equilibrium will all have about
the same $M_J$, as long as the galactic mass-to-light ratio is about
the same (expected variations in the IMF with M/L are discussed in
Elmegreen 1999c).

The transition from the power law part of the IMF to the low-mass
turnover should contain important information about the star formation
process, unlike the power-law part itself, which theory suggests
contains only information about hierarchical cloud structure, i.e.,
about the initial conditions for star formation.  The mathematical form
for this transition, i.e., the way in which the lower mass limit
actually affects the shape of the IMF, cannot be determined from theory
yet.  We have approximated this form simply by writing a probability
$P_f$ that a cloud piece fails to form a star on a time scale
comparable to its internal crossing time.  The functional form of this
probability is unknown (cf. eqn.  \ref{eq:imffun}), so the exact value
of the minimum critical mass cannot be known yet either. Differences of
a factor of three in the definition for $M_J$ can easily be compensated
by differences in the form of $P_f$. Nevertheless the concept that
thermal pressure in a star-forming cloud ultimately limits the mass of
a star is an essential part of the model, and leads to a characteristic
mass at the low end of the IMF that should scale with $T^2/P^{1/2}$.

The lower mass limit comes from another source in an alternative IMF
model, which follows from ideas by Larson (1982) and Shu, Adams, \&
Lisano (1987), namely, that protostellar winds limit the accretion onto
a star and thereby set the final stellar mass at a value close to the
lower limit necessary to drive such a wind (Nakano, Hasegawa, \& Norman
1995; Adams \& Fatuzzo 1996).  The wind-limited mass is not the
same as the deuterium-burning limit itself, which is about 
0.018 M$_\odot$ (D'Antona \& Mazzitelli 1994).
The lower stellar mass in this theory comes from a combination of the
deuterium limit and the accretion rate, $a^3/G$, for 
sound speed $a$ (Shu, Adams, \& Lisano 1987).
Higher accretion rates require stronger winds and more massive
stars before the accretion stops. This sensitivity to accretion
rate makes the lower mass limit depend on temperature in an
analogous way to $M_J$, so that lower temperature clouds are
likely to produce lower mass stars in the wind-limited
accretion model too.  The exact form of this temperature
dependence and whether there is a pressure
dependence are not known.

An important difference between these two models is that the
random-sampling model assumes that stellar mass is determined mostly by
the pre-stellar clump mass.  Then the IMF comes simply from random
selection of these clumps in a certain order (densest ones first).
There is, in fact, growing evidence that the stellar mass is determined
by the pre-stellar clump mass (Casoli et al.  1986; Myers, Ladd, \&
Fuller 1991; Myers \& Fuller 1993; Pound \& Blitz 1993, 1995; Motte, Andr\'e,
\& Neri 1998; Testi \& Sargent 1998), and not by wind-limited accretion
in a large reservoir.  The assumption of stellar mass proportional to
clump mass is reasonable if even a modest fraction, e.g., $>10$\%, of a
clump's mass goes into the stars it forms, because then the total
variability in possible stellar mass for a given clump mass is much
less than the total mass range of the IMF, which is a factor of 1000.
In this case most of the IMF has to come from clump mass variations and
not from variations in the star-to-clump mass fraction.  Moreover, even
if a random and small fraction of a clump's mass gets into a star and
the rest goes away in a wind or dispersed disk, the stellar mass will
still be proportional to the clump mass and the IMF will be unchanged
(Elmegreen 1999b) if the average of this fraction is about constant.

Given these differences between the two IMF models, the primary way to
distinguish between them is to determine the fraction of the clump mass
that goes into stars. If this fraction is either constant or relatively
large (larger than the inverse of the total mass range for stars), then
the IMF must be tracing the pre-stellar clump spectrum in some way. 

\section{A Prediction of Brown Dwarfs in Ultracold Gas}

IMF models with a minimum stellar mass that depends on the thermal
temperature and pressure in the star-forming cloud predict that
ultracold gas in the inner disk of M31 (Allen et al. 1995) should
produce a great abundance of Brown Dwarfs.  This gas is located in the
inner disk of M31, in the midst of the large, dark dust features
studied by Hodge (1980). The surface density of stars is comparable to
or higher there than it is locally because it is at a smaller
galactocentric radius than the Sun, and the surface density of the dark
CO clouds is higher than the local average too. Thus the pressure in
the cloud cores, and perhaps in the average interstellar medium at this
radius too, is larger than it is locally, by the product of the ratios
of these two surface densities.

This pressure change lowers $M_J$ by about a factor of 3 compared to
the value of $M_J$ for local star formation in molecular clouds.
Usually such a pressure increase is accompanied by a larger stellar
volume emissivity and cosmic ray flux in these inner disks, both of
which increase $T$ along with $P$. But in M31, the stellar blue surface
brightness and the non-thermal radio continuum flux are lower than in
the inner Milky Way, probably as a result of the earlier Hubble type
for M31. This implies that radiative and cosmic ray heating rates are
low, so the molecular gas equilibrates at a lower temperature, around
$3.5$ K (Allen et al. 1995; Loinard \& Allen 1998). In that case, there
is a factor of $\sim3$ drop in the temperature of potential
star-forming material for the ultracold CO gas in M31, and, when
combined with the expected higher pressure, a resulting net decrease in
$M_J$ by a factor of $\sim30$. This gives a value of \begin{equation}
M_J\sim0.01\;\;{\rm M}_\odot \end{equation} in the inner disk dark
clouds of M31.  If the pressure is the same in the cloud cores as it is
locally, then $M_J\sim0.03$ M$_\odot$.  Observation of a significant
population of $\sim0.01-0.1$ M$_\odot$ stars and Brown Dwarfs in the
inner dark clouds of M31 would therefore support theories of the IMF
based on either the thermal Jeans mass limit or the thermal accretion
rate limit for the lowest mass star.

The random sampling model also predicts that the slope of the IMF above
the thermal Jeans mass should be independent of $T$, $P$ and other
cloud properties since clouds presumably partition themselves in a
standard way by turbulence (for a discussion of cloud structure and
turbulence, see Falgarone, Phillips \& Walker 1991).  We expect the
same Salpeter slope, namely, $-1.35$ on a $\log-\log$ plot, for stars
with $M>>0.01$ M$_\odot$ in the M31 inner dark clouds as we find in the
intermediate to high mass parts of the IMF measured locally.  A
simulation showing this independence between the high-mass slope and
$M_J$ was shown by figure 4 in Elmegreen (1997).  Thus {\it the random
sampling model predicts that the IMF in ultracold clouds should be
exactly shifted towards lower mass without any change in the slope of
the power-law part.}

The observed IMF can be approximated by the function \begin{equation}
n_{log}(M)d\log M \approx A\left(1-e^{-\left(M/M_J\right)^\alpha}\right)
M^{-1.35}d\log M \label{eq:imffun} \end{equation} for constant $A$ and for
$\alpha$ in the range from 1 to 2; $\alpha\sim1$ for an IMF that
flattens at low mass, and $\alpha\ge2$ for an IMF that turns over at low
mass. There are only a few observations yet of a low mass turnover on a
$\log-\log$ plot (Reid \& Gazis 1997; Hillenbrand 1997; Nota et al. 1998), 
but numerous observations show a flattening. 
Note that there are usually many stars in this flattened region, and
that the minimum stellar mass can be as low as one-quarter the mass at
the turnover point.
In the
model, $\alpha$ appears only in the probability for failure to form a
star ($P_{f}\propto e^{-\left(M/M_J\right)^\alpha}$). 

Equation (\ref{eq:imffun}) can be integrated to give the total stellar
mass in various mass ranges. If $M_J=0.01$ M$_\odot$ and the Brown
Dwarf limit is $M_{BD}\sim0.08$ M$_\odot$, then 0.479 of all the mass
$M>M_J$ will be between $M_J$ and $M_{BD}$.  {\it Approximately half the
stellar-like mass down to the peak in the IMF would be in the form of
Brown Dwarfs.}

\section{Observable consequences of a Brown Dwarf bias in ultracold
star formation}

The formation of a sufficiently large density of Brown Dwarfs in an
ultracold CO cloud could raise the gas temperature and make
it more ``normal.'' The energy input required to heat a molecular cloud
to $\sim10$K is about $10^{-26.6}n_0$ erg cm$^{-3}$ s$^{-1}$ for
molecular density $n_0$ (Neufeld, Lepp \& Melnick 1995). This
corresponds to a ratio of luminosity to mass equal to $6.4\times10^{-4}$
erg s$^{-1}$ g$^{-1}$. If the ratio of the stellar luminosity that heats
the gas divided by the gas mass, $\left(L/M\right)_y$, from Brown Dwarfs
and other young stars is close to this, then the gas would not be
supercold. A typical $L/M$ for pre-main sequence Brown Dwarfs is $\sim1$
in these units (D'Antona \& Mazzitelli 1994), so if the star formation
efficiency is 1\% and a high fraction of the stellar radiation heats the
gas, then the stellar luminosity per unit total mass would be 0.01 erg
s$^{-1}$ g$^{-1}$, which is 16 times larger than the ratio needed to
heat the cloud to 10K. 

Most embedded stellar radiation does not go into the gas directly,
though, it goes into the dust which radiates it away in the IR. The
dust luminosity scales approximately with dust temperature as $T^5$
(e.g., Hollenbach \& McKee 1979), and a typical cold dust temperature
in M31 is $\sim16$ K (Haas et al. 1998). If the dust and gas are
thermally coupled in ultracold CO clouds, then the dust luminosity has
to be less in $\sim3$K clouds than it is locally by a factor of about
$5^5\sim3000$.  The background radiation field in M31 is brighter than
$1/3000$ times the local value, so the dust and gas cannot be well
coupled in the M31 ultracold CO clouds. This means that the average gas
density in these clouds is much less than $\sim10^4$ cm$^{-3}$.  The
density will be higher than this in star-forming cores, but these are
shielded from outside light by dust, so the core dust temperature
should be low there.

The likely range for the average cloud density can be estimated from the
inclination-corrected column density of $\sim100$ M$_\odot$ pc$^{-2}$
(Loinard \& Allen 1998). Considering a typical cloud projected size of
$\sim100$ pc, the cloud thicknesses are probably somewhere between
$\sim1$ pc if they are thin shells, and $\sim100$ pc, the thickness of
the galaxy. This puts the average cloud density between 1 and 100
M$_\odot$ pc$^{-3}$, which corresponds to a molecular hydrogen density
of 15--1500 cm$^{-3}$. This is consistent with the result of Loinard,
Allen \& Lequeux (1995), who estimated a density of $\sim100$ cm$^{-3}$
from CO line ratios. Thus the average density is indeed too low for
thermal coupling between the gas and dust.

A better way to determine if embedded Brown Dwarfs can significantly
heat ultracold CO is to compare the summed luminosities of these stars
to the incident and embedded luminosity from field stars. If the Brown
Dwarf luminosity in the clouds is less than the total absorbed field
star power, then the clouds would not show any excess emission in either
dust or gas from the embedded young stars. 

The total field star power received by a spherical cloud of radius $R$
is \begin{equation} P_f=4\pi R^2 j_f\left(\pi
\lambda+R/3\right)\end{equation} for field star volume emissivity $j_f$
and average pathlength $\lambda$ for field star radiation. The first
contribution in the parenthesis is from external field stars, and the
second is from internal field stars. The total luminosity of embedded
young stars with volume emissivity $j_y$ is \begin{equation}
L_y=\left(4/3\right) \pi R^3j_y.\end{equation} The ratio $L_y/P_f$ has
to be large for the cloud temperature to increase significantly as a
result of embedded star heating. This ratio gives a critical ratio of
volume emissivities \begin{equation} {{j_y}\over{j_f}}>1+{{3\pi
\lambda}\over{R}}\sim10^2-10^3 \end{equation} for embedded star heating;
here we have taken an external path length equal to $10-100$ times the
cloud size, considering this as the ratio of cloud mean free path to
size, or the inverse of the volume filling factor of these ultracold CO
clouds. 

The volume emissivity of embedded stars is
$j_y=\left(L/M\right)_y\epsilon \rho$ for average stellar
luminosity-to-mass ratio $\left(L/M\right)_y$, star formation
efficiency $\epsilon$ (the ratio of the star mass to the total cloud
mass), and gas density $\rho$. To find the average $\left(L/M\right)_y$
as a function of time, we integrate the pre-main sequence stellar
luminosities over the stellar mass function, \begin{equation}
\left({{L}\over{M}}\right)_y(t)={{\int_{M_J}^{M_U} L(M,t)n(M)dM}\over
{\int_{M_J}^{M_U} Mn(M)dM}} \label{eq:lmburst} \end{equation} using the
Alexander + Rogers \& Iglesias, CM model for $L(M,t)$ in D'Antona \&
Mazzitelli (1994) and $n(M)dM=n_{log}(M)d\log M$ from equation
\ref{eq:imffun}. These pre-main sequence models are only for $M<M_U=2.5$
M$_\odot$, so our results are valid only in this limit. Equation
\ref{eq:lmburst} gives $\left(L/M\right)_y$ for a time $t$ after a short
burst of star formation. To get $\left(L/M\right)_y$ from continuous
star formation that has lasted for a time $t$, we use: \begin{equation}
\left({{L}\over{M}}\right)_{y,cont}(t)={{\int_0^t dt \int_{M_J}^{M_U}
L(M,t){\cal R}(M)dM}\over {\int_0^t dt \int_{M_J}^{M_U} M{\cal R}(M)dM}}
\end{equation} for star formation rate by number, ${\cal R}(M)$, which
is proportional to $n(M)$ for a uniform star formation rate. 
The lower limits to these integrals are taken to be $M_J$ instead of
the absolute minimum stellar mass, because the form of the IMF is not
well known below the peak and because lower mass Brown Dwarfs
should not contribute much to the luminosity anyway.

The burst
and continuous $\left(L/M\right)_y$ are shown in Figure 1 as functions
of time since the burst and as functions of age, respectively, for
$M_J=0.01$ M$_\odot$, and for mass ranges 0.02--1 and 0.02--2.5
M$_\odot$. For continuous star formation, $\left(L/M\right)_y\sim2$
L$_\odot$/ M$_\odot\sim4$ erg s$^{-1}$ g$^{-1}$ for this IMF. For the
larger mass range, the burst $\left(L/M\right)_y$ levels off because
massive stars dominate the light and reach the main sequence after
$\sim10^6$ years. 

The volume emissivity of field stars is $j_f=\left(L/M\right)_f\rho_f$.
For Sb galaxies like M31, $\left(L/M\right)_f\sim0.5$
L$_\odot$/M$_\odot$ (Roberts \& Haynes 1994). Thus the condition for
significant young star heating is \begin{equation}{{\rho}\over{\rho_f}}
> {{\left(10^2-10^3\right)}\over{\epsilon}}{{\left(L/M\right)_f}\over
{\left(L/M\right)_y}}\sim{{25-250}\over{\epsilon}}. \label{eq:rhorho}
\end{equation} The background stellar density in the inner M31 disk is
several M$_\odot$ pc$^{-3}$, and the density in the ultracold CO clouds
is in the range 1--100 M$_\odot$ pc$^{-3}$, given above, so
$\rho/\rho_f\sim$ 0.5--50. This is comparable to the critical density
ratio for young star heating only if $\epsilon>>0.1$. 

{\it We conclude that a Brown Dwarf+stellar population with $M<2.5$
M$_\odot$ would not heat the ultracold CO clouds in M31 noticeably if
the star-to-cloud mass ratio is less than 10\%.}

Another way to get a limit on how many Brown Dwarfs and other young
stars might be present is from a limit on the mass of the most massive
star that can be associated with these clouds. The largest stellar mass,
$M_{max}$, that is likely to come from an IMF with a total stellar mass
$M_{tot}$ is given by \begin{equation} {{\int_{M_J}^\infty
Mn(M)dM}\over{\int_{M_{max}}^\infty n(M)dM}}=M_{tot}. \end{equation}
Figure 2 (solid lines, left axis) shows the total stellar and Brown
dwarf masses versus $M_{max}$ (the two lines nearly overlap). If the
largest pre-main sequence or stellar mass that can be hidden in and
around the ultracold CO clouds in M31 is $\sim2$ M$_\odot$, then the
total mass of all the Brown Dwarfs+stars has to be less than $\sim 50$
M$_\odot$. Figure 2 (dashed lines, right axis) shows the numbers of
Brown Dwarfs (upper dashed line) and stars as functions of $M_{max}$.
With $M_J\sim0.01$ M$_\odot$, {\it there could be $\sim10^3$ Brown
Dwarfs and $\sim80$ stars less massive than 2.5 $M_\odot$ in each of
the large ultracold CO clouds in M31}.

\section{A low star formation rate}

This number of $\sim10^3$ hidden Brown Dwarfs in ultracold CO clouds may
seem large compared to what has been found so far in the Solar
neighborhood, but the associated number of normal stars and the total
mass in the form of stars or Brown Dwarfs is remarkably small for such a
cloud with an estimated $\sim10^6$ M$_\odot$ of gas (Loinard \& Allen
1998). If the efficiency of star+Brown Dwarf formation is anything like
it is locally, namely $\ge1$\%, then so many stars should have formed
that the IMF would have sampled out far enough into the high mass tail
to produce O-type stars, which would be seen easily. This implies that
star formation is unusually inefficient in the M31 ultracold CO clouds.

One possible explanation for this is that the clouds are much less dense
than the excitation densities given by Loinard \& Allen (1998). These
authors found, on the basis of excited-state CO line ratios, that the
gas is optically thick in the 3-2, 2-1, and 1-0 $^{12}$CO lines. In that
case, the density required for excitation is less by the factor $1/\tau$
for opacity $\tau$ than the density that makes the collision rate equal
to the spontaneous transition rate $A$. Thus the molecular density can
be $\sim10$ times less than their estimate, and the clouds much less
strongly self-gravitating. If the clouds are like galactic translucent clouds,
they may form no stars at all (e.g., Hearty et al. 1999).

We can use the example provided by M31 to estimate the importance of
Brown Dwarf formation generally. Suppose the efficiency of Brown
Dwarf+star formation in ultracold gas is the upper limit given by the
M31 clouds, which is $\sim10^{-4}$ in $10^6$ M$_\odot$ clouds to avoid
stars more massive than 2 M$_\odot$. Suppose also that half the
stellar-like mass goes into Brown Dwarfs and that the ages of the ultracold
clouds are at least $\sim10^7$ years, which is a modest fraction of the
shear time in the inner disk M31. Then the Brown Dwarf formation rate is
$<5$ M$_\odot$ My$^{-1}$ per $10^6$ M$_\odot$ of ultracold gas. This
implies a gas consumption time that is very large, $2\times10^{11}$
years, which means hardly any conversion of gas into stars or
Brown Dwarfs in a Hubble
time. To get a significant mass in Brown Dwarfs, the efficiency of star
formation has to be larger than this by a factor of $\sim100$, but then
the resulting luminosity would heat an ultracold cloud to normal
temperatures, shutting off Brown Dwarf production in favor of regular
stars. 

The formation of Brown Dwarfs in ultracold gas requires a negligible
mass of normal stars so the gas remains cold. This would occur naturally
in the lowest mass clouds from an ensemble of clouds if the stars
formed randomly. For example, if the maximum total mass that can form
in a cloud before a massive star is likely to heat the gas is $\sim100$
M$_\odot$, from figure 2, and if the efficiency is large, 10\%, then
the clouds that make Brown Dwarfs have to be smaller than $10^3$
M$_\odot$.  Such small clouds are rarely isolated, however, and nearby
gas can still make massive stars that provide heat.  Thus is it
unlikely Brown Dwarfs can form with a significant total mass if the IMF
is simply shifted towards a lower peak mass with a Salpeter spectrum
above this. An upper mass cutoff in star formation, which is not
present in the theory, would seem to be necessary if Brown Dwarfs are
ever found to dominate the mass in a region.

\section{Conclusions}

The ultracold clouds in M31 provide a good test for theories of the
IMF.  Because of their low temperatures and normal-to-high pressures,
these clouds should have a thermal Jeans mass of only $\sim0.01$
M$_\odot$, making the IMF shift towards lower mass with the same slope
at higher mass. In that case, ultracold clouds should produce half of
their stellar-like mass in the form of Brown Dwarfs. If the total
number of such Dwarfs per cloud is $\le10^3$, then they and their
accompanying H-burning stars would not significantly heat the cloud,
nor be visible in existing surveys. A deep K-band search for Brown
Dwarfs in ultracold gas would be necessary to see them.  Considering
the luminosities of such pre-main sequence stars found locally, which
is $K\sim13$ mag dereddened (Luhman et al. 1998), the K band luminosity
of such a star in M31 would be $\sim 21$ mag plus extinction.

Helpful comments by the referee are appreciated.

\newpage
\begin{figure}
\vspace{3.in}
\includegraphics{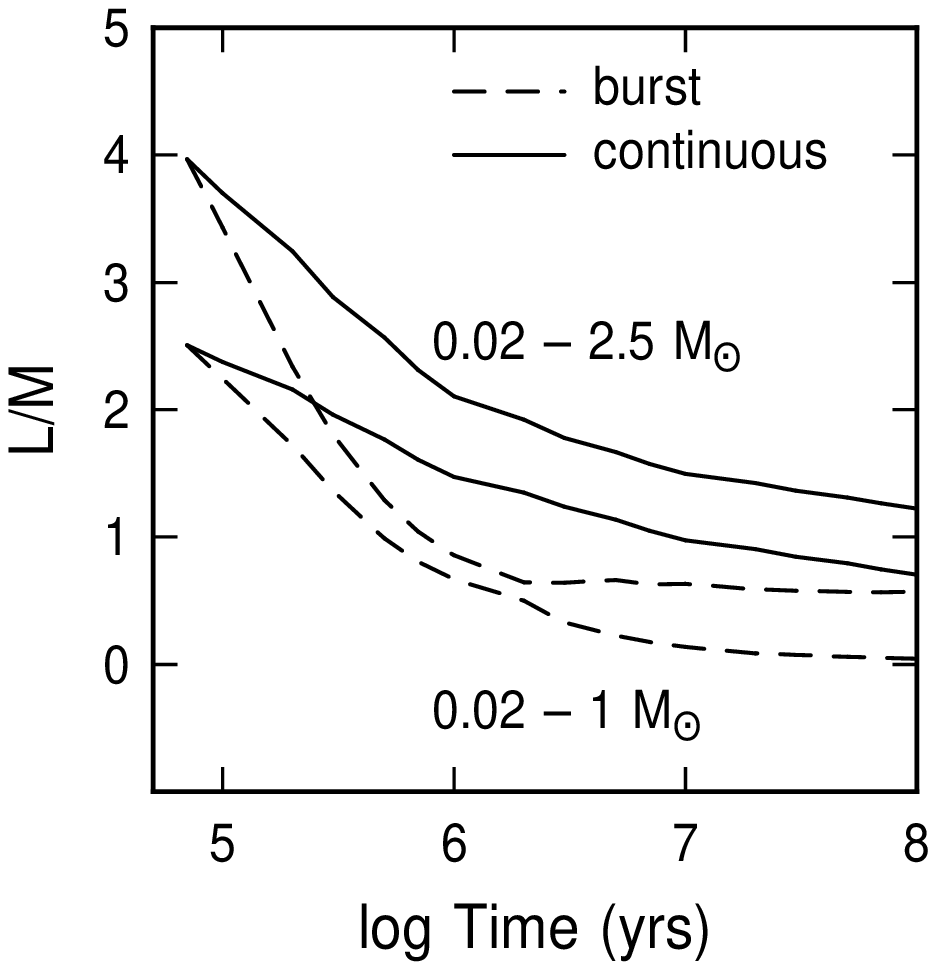}
\caption{The luminosity to mass ratio, in units of L$_\odot$/M$_\odot$, from
pre-main sequence and main-sequence stars in the mass ranges indicated,
are plotted as functions of time for a burst at $t=0$ and for continuous
star formation. These $L/M$ ratios assume an initial mass function
of the form given by equation (3) with $M_J=0.01$ M$_\odot$. }
\label{fig:loverm}
\end{figure}

\begin{figure}
\vspace{3.in}
\includegraphics{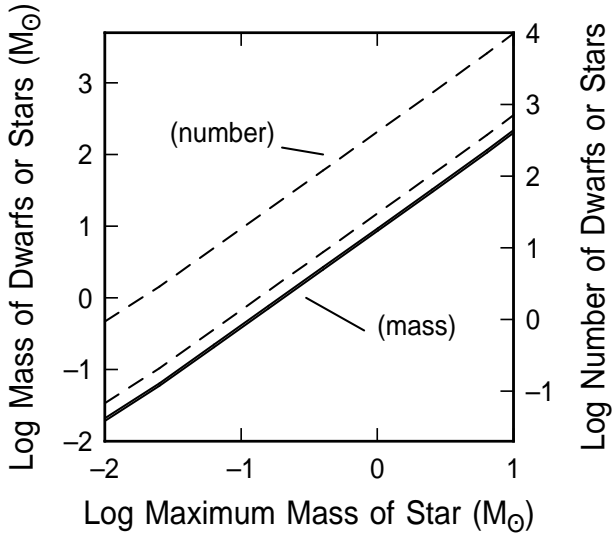}
\caption{The total stellar and Brown Dwarf masses are shown with the
left hand axis as solid lines (these masses are
approximately equal, so the two lines are nearly superposed),
and the cluster stellar and Brown Dwarf numbers are
shown with the right-hand axis, as dashed lines.  
The values assume $M_J=0.01$ M$_\odot$.  The Brown Dwarf
dashed line is above the star dashed line. Both are plotted 
as functions of the mass of the
largest star that is likely to form.}
\label{fig:mm}
\end{figure}

\end{document}